\documentclass[%
 reprint,%
 amssymb, amsmath,%
aip 
]{revtex4-1}

\usepackage{graphics}
\usepackage{amsmath}
\usepackage{amsfonts}
\usepackage{latexsym}
\usepackage{amsbsy}
\usepackage{epstopdf}
\usepackage{bm}%

\usepackage{color}

\usepackage{ulem} 

\newcommand{\dyadic}[1]{{#1}
\setbox0=\hbox{$\scriptstyle\leftrightarrow$}
   \setbox2=\hbox{$#1$}
   \dimen0=.5\wd0 \advance\dimen0 by-.5\wd2
   \advance\dimen0 by-\wd0
   \kern\dimen0
{^{\hbox{$\scriptstyle\leftrightarrow$}}}}

\begin{document}

\title{
{Rydberg atom-based field sensing enhancement using a split-ring resonator}\\
}
\thanks{Publication of the U.S. government, not subject to U.S. copyright.}
\author{Christopher L. Holloway}
\email{christopher.holloway@nist.gov}
\author{Nikunjkumar Prajapati}
\author{Alexandra B. Artusio-Glimpse}
\author{Samuel Berweger}
\author{Matthew T. Simons}
\affiliation{National Institute of Standards and Technology, Boulder,~CO~80305, USA}
\author{Yoshiaki Kasahara}
\affiliation{The University of Texas, Austin, TX,~USA}
\author{Andrea Alu}
\affiliation{Advanced Science Research Center, City University of New York, NY, USA}
\author{Richard W. Ziolkowski}
\affiliation{University of Technology Sydney, Ultimo NSW, Australia
}
\date{\today}

\begin{abstract}
We investigate the use of a split-ring resonator (SRR) incorporated with an atomic-vapor cell to improve the sensitivity and the minimal detectable electric (E) field of Rydberg atom-based sensors. In this approach, a sub-wavelength SRR is placed around an atomic vapor-cell filled with cesium atoms for E-field measurements at 1.3~GHz. The SRR provides a factor of 100 in the enhancement of the E-field measurement sensitivity. Using electromagnetically induced transparency (EIT) with Aulter-Townes splitting, E-field measurements down to 5~mV/m  are demonstrated with the SRR, while in the absence of the SRR, the minimal detectable field is 500~mV/m.  We demonstrate that by combining EIT with a heterodyne Rydberg atom-based mixer approach, the SRR allows for the a sensitivity of 5.5~$\mu$V/m$\sqrt{{\rm Hz}}$, which is two-orders of magnitude improvement in sensitivity than when the SRR is not used. 

\end{abstract}

\maketitle


In the past ten years, great progress has been made in the development of Rydberg atom (atoms with one or more electrons excited to a very high principal quantum number $n$)\cite{gal} based radio-frequency (RF) electric (E) field sensors. The majority of this work relies on spectroscopy of highly excited atoms using electromagnetically induced transparency (EIT) and their interaction with external electric fields in the form of Autler-Townes (AT) splitting\cite{sed1, holl1, holl2}. Other measurements involving non-resonant Stark shifts have also been performed. These sensors now have the capability of measuring the amplitude \cite{gor1, sed1, holl1, holl2, tan1, kumar1, r5, gor2, gor3}, polarization \cite{sed2, access}, and phase \cite{sim3, jing1, dphase} of the RF field, and various applications are beginning to emerge. These include E-field probes traceable to the International System of units (SI)\cite{holl1, holl2, gor2}, power-sensors \cite{holl5}, spectrum analyzers \cite{army}, voltage standards\cite{volt}, angle-of-arrival detection\cite{aoa}, and receivers for communication signals (AM/FM modulated and digital phase modulation signals) \cite{song1, meyer1, holl6, cox1, holl4, anderson2, deb3, holl7}.


\begin{figure}[t!]
\centering
\scalebox{.40}{\includegraphics*{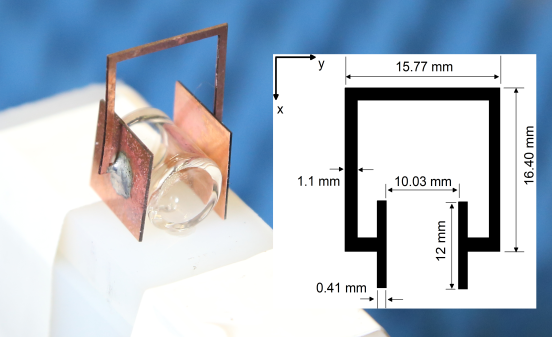}}\\
\caption{Vapor cell place inside the split-ring resonator (SRR). The insert shows the dimension of the SRR.}
\label{srr}
\end{figure}

A current thrust in the development of atom-based sensors is geared towards improving the minimal detectable field and sensitivity. Using standard EIT/AT techniques, E-field strengths down to a few V/m can routinely be measured. Even lower fields, down to tenths of V/m, can be measured depending on the frequency and atomic states used\cite{sed1, holl1, holl2, r15}. One issue with using standard EIT/AT approaches for weak field detection stems from the limitations on EIT/AT based spectroscopy. Measuring a very weak RF field requires the measurement of very small AT splitting (or small Rabi frequency). The AT splitting (in units of Hz) is directly related to the Rabi frequency ($\Omega_{RF}$) of the applied RF field\cite{sed1, holl1}:
\begin{equation}
{\rm AT\,\,splitting}=\frac{\Omega_{RF}}{2\pi}=|E| \frac{\wp}{h}\,\,\,\, ({\rm Hz})
\label{e1}
\end{equation}
where $|E|$ is the magnitude of the E-field, $\wp$ is the
atomic dipole moment, and $h$ is Planck's constant. The measurement of the AT splitting and, therefore, field measurement, is limited by the resolution set by the EIT linewidth, which is typically on the order of 2 MHz to 5~MHz. This makes AT splittings smaller than this linewidth difficult to determine.  The primary factor driving the broadening of the line is the Doppler mismatch of the probe and coupling lasers used in this two-photon approach\cite{r15}.
The EIT linewidth is one of the limiting factors that affect the minimum detectable E-field and the sensitivity of the Rydberg atom sensor. In fact, it is shown in in Ref.\cite{r15} that reliable and accurate SI-traceable measurements are only possible when the $\Omega_{RF}$ is greater than twice the EIT linewidth.


There are approaches to overcome the Doppler mismatch and hence improve minimal weak field detection. One such approach is to use three-photon excitation schemes to reach the Rydberg state \cite{shaffer2018, shaffer2021, carr2012three, Adams2019}. While there are groups researching the three-photon scheme, weak field detection with this approach has not been demonstrated yet. Here, we take an alternative approach. We discuss the use of a split-ring resonator (SRR) structure, see Fig.~\ref{srr}, to enhance the incident E-field at the location of the atoms inside the vapor cell, hence improving the sensitivity of the Rydberg-atom sensor. The use of metallic structures to enhance the sensitivity or the polarization selectively have been used in the past by either placing parallel plate structures inside the vapor cell\mbox{\cite{david}} or by embedding the vapor cell inside waveguiding structures\mbox{\cite{access}}. In this paper, we demonstrate that a sub-wavelength resonator structure can be placed around a atomic-vapor cell to improve the sensitivity of Rydberg atom sensors by a factor of 100.


It has been shown that sub-wavelength SRRs can substantially enhance an incident field in a confined region in the gaps of the SRR, 
see Fig.~7 of Ref.\cite{josh}. To exploit this, we place a SRR around a 10.03~mm outside-diameter vapor cell, see Fig.~\ref{srr}.  In this type of structure, the ring captures the incident RF field and enhances it, i.e., the field induced between the gap is larger than the incident field. Thus, when a vapor-cell is placed in the gap, the atoms will be exposed to a field that is larger than the incident field. The increase of the field in the vapor cell will be the enhancement factor. 
The loop size and gap separation of the SRR controls the resonance frequency\cite{syoruk}. In our design we were limited to a gap separation of 10.03~mm, because of the available size of the vapor cell (the outside diameter of the cell). We used a commercial finite-element simulator to determine the loop size for a SRR to have a resonant frequency near 1.312~GHz. This frequency corresponds to a Rydberg atom transition for the states $80D_{5/2}$ and $81P_{3/2}$. The final design is shown in Fig.~\ref{srr} and is made from copper of thickness 0.41~mm and the loop has side lengths of 15.77~mm by 16.40~mm and width of 1.10~mm. The plates are 12~mm by 12~mm and separated by 10.03~mm. 

\begin{figure}[t!]
\centering
\scalebox{.39}{\includegraphics*{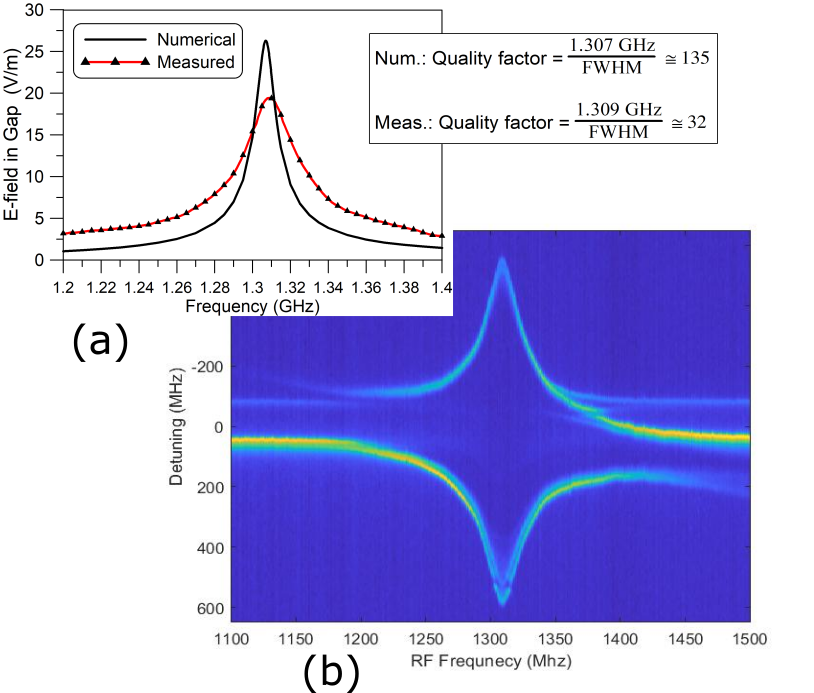}}\\
\vspace{-3mm}
\caption{Frequency response of the SRR: (a) Electrical field in gap of SRR. Comparison of the numerically calculation and the measured E-field, and (b) contour plots of the EIT signal for RF frequency detuning showing AT splitting. Both results are for a SG power of -14~dBm (or a 0.2~V/m incident E-field). }
\label{compare}
\end{figure}

\begin{figure}
\centering
\scalebox{.067}{\includegraphics*{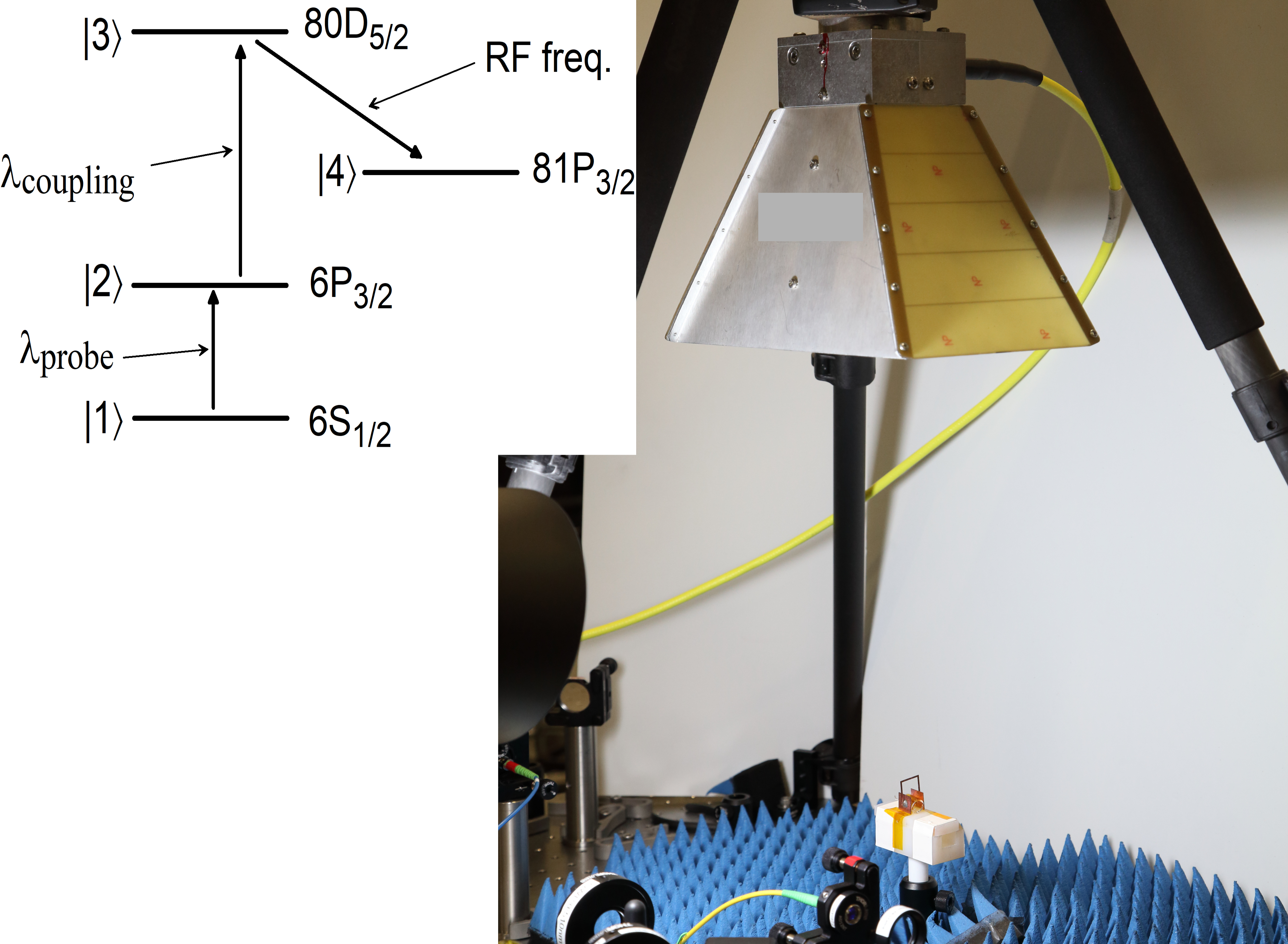}}\\
\vspace{-3mm}
\caption{Experimental setup for the SRR experiment. The insert shows the atomic levels used in the EIT scheme.}
\label{setup}
\end{figure}

The numerically calculated E-field values between the SRR gap are shown in Fig.~\ref{compare}(a). These numerical results include the dielectric amounting structure seen in Figs.~\ref{srr} and \ref{setup}. We see that this modeled structure has a resonance at 1.307~GHz and has fairly narrow frequency response, which is typical of resonant structures. Also shown in this figure are the measured field values in the gap (a discussion on how these measurements are performed is given below). 
While we see a good correlation between the modeled and measured data, there is a slight shift in the measured resonator frequency to 1.309~GHz. Also, the measured peak structure is slightly squashed and broadened compared to the model. The measured enhancement is roughly 100 compared to the modeled result of 130. This discrepancy is likely due to losses (metallic and dielectric) in the SRR structure, and likely variances in dimensions and fabrication.
Furthermore, we see that the measured resonance, 1.309~GHz, is only 0.2~\% from the atomic resonance for the 80D$_{5/2}$-to-81P$_{3/2}$ states at 1.312~GHz. While off resonance RF frequencies can effect these types of EIT/AT measurements\cite{detuning}, our low 0.2~\% shift from the atomic resonance will have little effect on our measurements. Beside being frequency selective, this type of SRR is highly polarization selective. This design enhances an incident plane-wave field when the E-field is polarized in the $y$-direction (see Fig.~\ref{srr}) and the magnetic field is polarized along the $z$-axis. 


The experimental setup is depicted in Fig.~\ref{setup}. It consists of a 852~nm probe laser, a 508~nm coupling laser, a horn antenna, two photodetectors connected to an oscilloscope, and the vapor cell shown in Fig.~\ref{srr} filled with cesium ($^{133}$Cs) atomic vapor. The vapor cell is a cylindrical cell 14~mm long, and a outside diameter of 10.03~mm (the thickness of the cell walls are 1~mm).  We use a three-level EIT scheme to generate Rydberg atoms (see insert in Fig.~\ref{setup}) which correspond to the $^{133}$Cs  $6S_{1/2}$ ground state,  $6P_{3/2}$ excited state, and a Rydberg state 80$D_{5/2}$ state. We use a 1.3~GHz RF source to couple to the  81$P_{3/2}$ state.

The probe laser is locked to the D2 transition (\mbox{$6S_{1/2}(F=3)$~--~$6P_{3/2}(F=4)$} whose wavelength is \mbox{$\lambda_p=852.35$~nm} \cite{stackcs}).  To produce an EIT signal, we apply a counter-propagating coupling laser with $\lambda_c \approx 508$~nm and scan it across the $6P_{3/2}-80D_{5/2}$ Rydberg transition. We use two photodetectors in differential detection and a lock-in amplifier to enhance the EIT signal-to-noise ratio by modulating the coupling laser amplitude with a 37~kHz square wave. This removes the background and isolates the EIT signal.  In these experiments, the optical beams and the RF electric fields  are co-linearly polarized. The probe laser was focused to a full-width at half maximum (FWHM) of  800~$\mu$m with a power of 88~$\mu$W, and the coupling laser was focused to a FWHM of  300~$\mu$m with a power of 31~mW. A signal generator (SG) is connected to the horn antenna and the output power of the SG is varied during the experiments. The antenna face is 310~mm from the vapor cell.

\begin{figure}
\centering
\scalebox{.33}{\includegraphics*{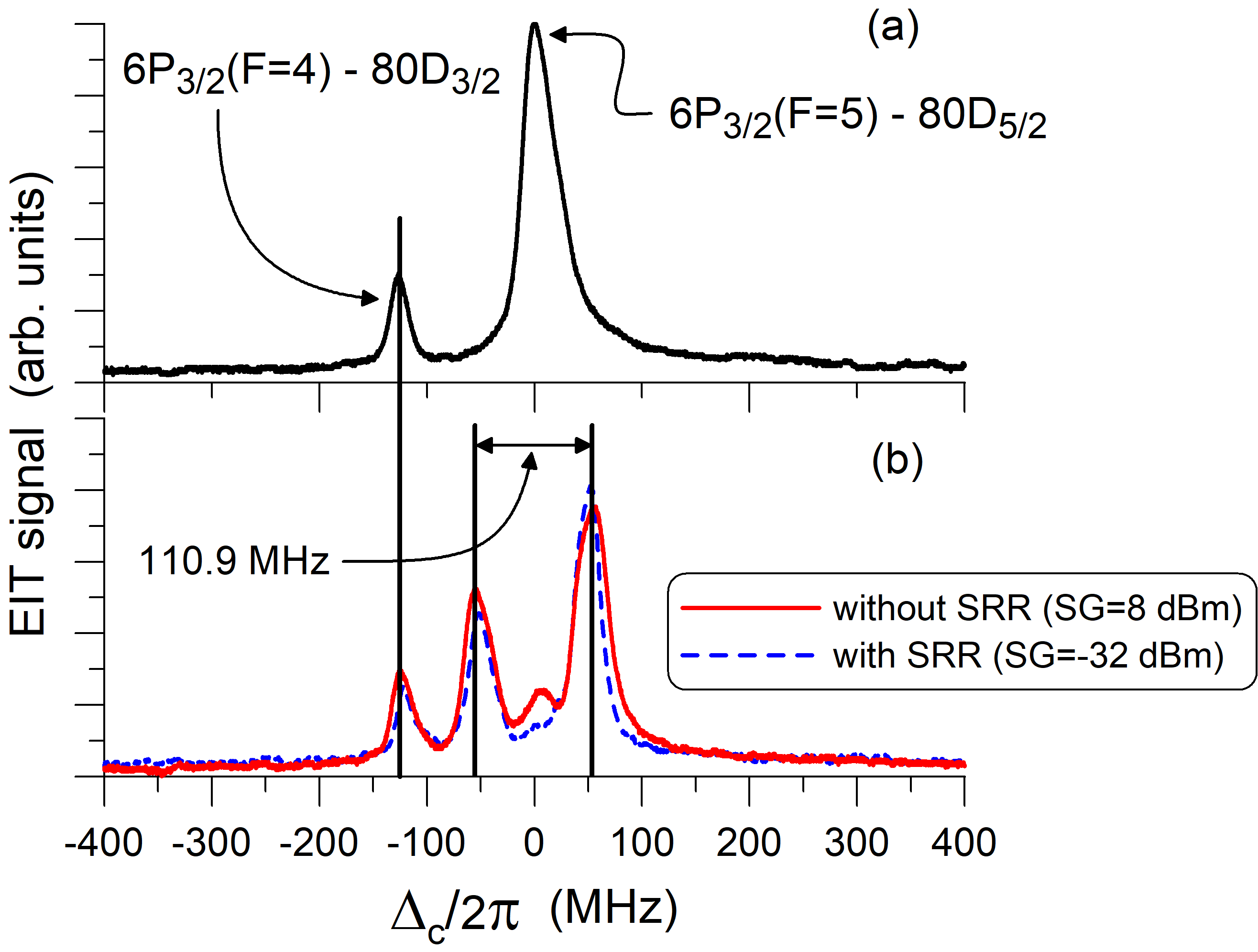}}\\
\vspace{-3mm}
\caption{Measured EIT signal versus coupling laser detuning ($\Delta_c$): (a) no RF field and (b) AT splitting without the SRR for a SG power of 8~dBm and with the SRR for a SG power of -32~dBm.}
\label{eitsignal}
\end{figure}

Fig.~\ref{eitsignal} shows three EIT signals as functions of the coupling laser detuning ($\Delta_c$). The top curve is with no RF field (only the probe and coupling lasers). The two different peaks correspond to the transitions from $6S_{1/2}$ to two allowed $6P_{3/2}$ fine-structure levels ($80D_{5/2}$ and $80D_{3/2}$). The larger main peak at $\Delta_c/2\pi=0$ corresponds to the transition \mbox{$6P_{3/2}$(F=5)$-80D_{5/2}$}. When a 1.309~GHz RF field is applied, this peak will experience AT splitting. The second peak at $\Delta_c/2\pi$=-129.1~MHz corresponds to the transition \mbox{$6P_{3/2}$(F=5)$-80D_{3/2}$}. This peak separation is used to calibrate the frequency axis, where the time axis from the scope is converted to frequency by noting that the peaks are 129.1~MHz apart. When a 1.309~GHz RF field is applied, this peak will not experience AT splitting as we will see in the data sets below.

A 1.309~GHz RF source couples the two Rydberg states 80$D_{5/2}$ and 81$P_{3/2}$. If the field strength is large enough, this 1.309~GHz field will induce AT splitting in the main EIT line.  The solid curve in Fig.~\ref{eitsignal}(b) is for the case without the SRR when the RF source is turned on with the output of the SG set to 8~dBm. The dashed curve in Fig.~\ref{eitsignal}(b) is for the case with the SRR when the RF source is on with the output of the SG set to -32~dBm.

Without the SRR, the 8~dBm SG output power causes an AT splitting of 110.9~MHz, while with the SRR, the same AT splitting is caused by a SG output power of \mbox{-32~dBm}.  Note that without the SRR, no AT slitting is observed in the EIT signal for \mbox{-32~dBm}. Thus, 40~dB less power is required to detect the same AT splitting (or the same E-field inside the vapor cell). This indicates that the SRR enhanced the field by a factor $\sqrt{10^{40/10}}$=100. This demonstrates that the atomic sensor utilizing the SRR can detect E-fields that are 100 times weaker than a sensor without the SRR.

By plotting the EIT signal for different RF frequencies, the resonance frequency of the SRR structure can be determined. Contour plots for the EIT signal, when the RF frequency is varied, are shown in Fig.~\ref{compare}(b). The contour plots are composed of several EIT traces. We see the AT splitting increases as the RF frequency is increased  and reaches a maximum at 1.309~GHz. As the frequency is further increased, past 1.309~GHz, the AT splitting begins to decrease. Using these AT splittings and Eq.~(\ref{e1}), the E-field between the gaps of the SRR structure is determined. These measured E-field values are shown (and compared to numerically modeled values) in Fig.~\ref{compare}(a). These values are for a SG power of -14~dBm, which corresponds to an 0.2~V/m at the vapor cell (this E-field value is determined with the SG power level and Eq.~(3) in Ref.\cite{six}).  The uncertainties of these types of measurements are related to the EIT/AT detection scheme and are on the order of 1~\%\cite{sed1, r15, Simons2018}.

\begin{figure}
\centering
\scalebox{.33}{\includegraphics*{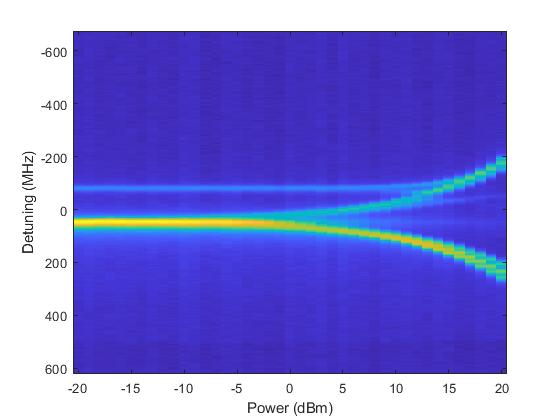}}\\
\scalebox{.33}{\includegraphics*{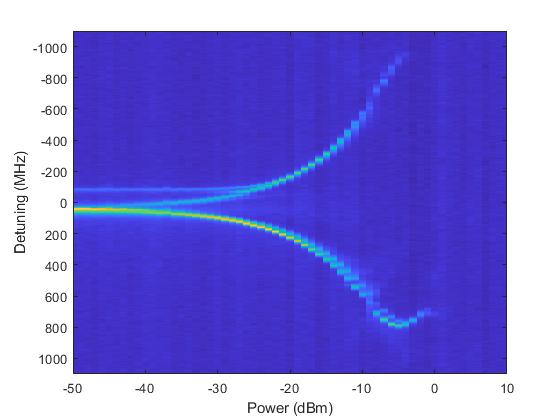}}\\
\vspace{-3mm}
\caption{Contour plot of a series of EIT signals as a function of the SG power and coupling detuning ($\Delta_c$): (a) without the SRR and (b) with the SRR.}
\label{contour}
\end{figure}

To further show the field enhancement (or field concentration) facilitated by the SRR, we measured the EIT signal for a range of SG output power levels. Fig.~\ref{contour} shows the contour plots of the EIT signal as a function of the SG output level with and without the SRR. Fig.~\ref{contour}(a) is the case without the SRR. We see that the AT splitting starts to appear around -3~dBm. Fig.~\ref{contour}(b) is the case with the SRR. We see that the AT splitting starts to appear around -43~dBm.  Also, in both figures, we see that the EIT line at 129.1~MHz (the 80D$_{3/2}$ state) does not split with increasing SG power levels.  In Fig.~\ref{contour} we see that a more complicated structure in the spectra is observed with the SRR when compared to the no SRR results. For example, in Fig.~\ref{contour}(b) we see a curving upward of the lower EIT line of about -5~dBm.  This is due to the higher field that occurs at the atoms when the SRR is used.  Note that the SG used can only output a maximum of 20~dBm. This added structure is a result of the much higher E-field at the location of the atoms due to the SRR, in that the atomic spectra becomes highly nonlinear in this strong field regime.  This type of feature is observed in strong E-field spectrometry \cite{anderson2016} and is a result of state mixing and higher-order couplings. As stated in Ref.\cite{anderson2016}, these features allow for precision measurements of strong E-field strengths. The SRR enhancement causes the highly nonlinear features. These features are not observed without the SRR because of the limited output power of the SG, but the SRR enhances the field at the location of atoms by a factor of 100 (or 40~dB). 
Without the SRR the EIT/AT approach can detect a field level around 500~mV/m (the point where AT slitting can first be observed). On the other hand, and with the SRR, we can detect a field down to 5~mV/m, an enhancement factor of 100 in field strength detection.

\begin{figure}
\centering
\scalebox{.30}{\includegraphics*{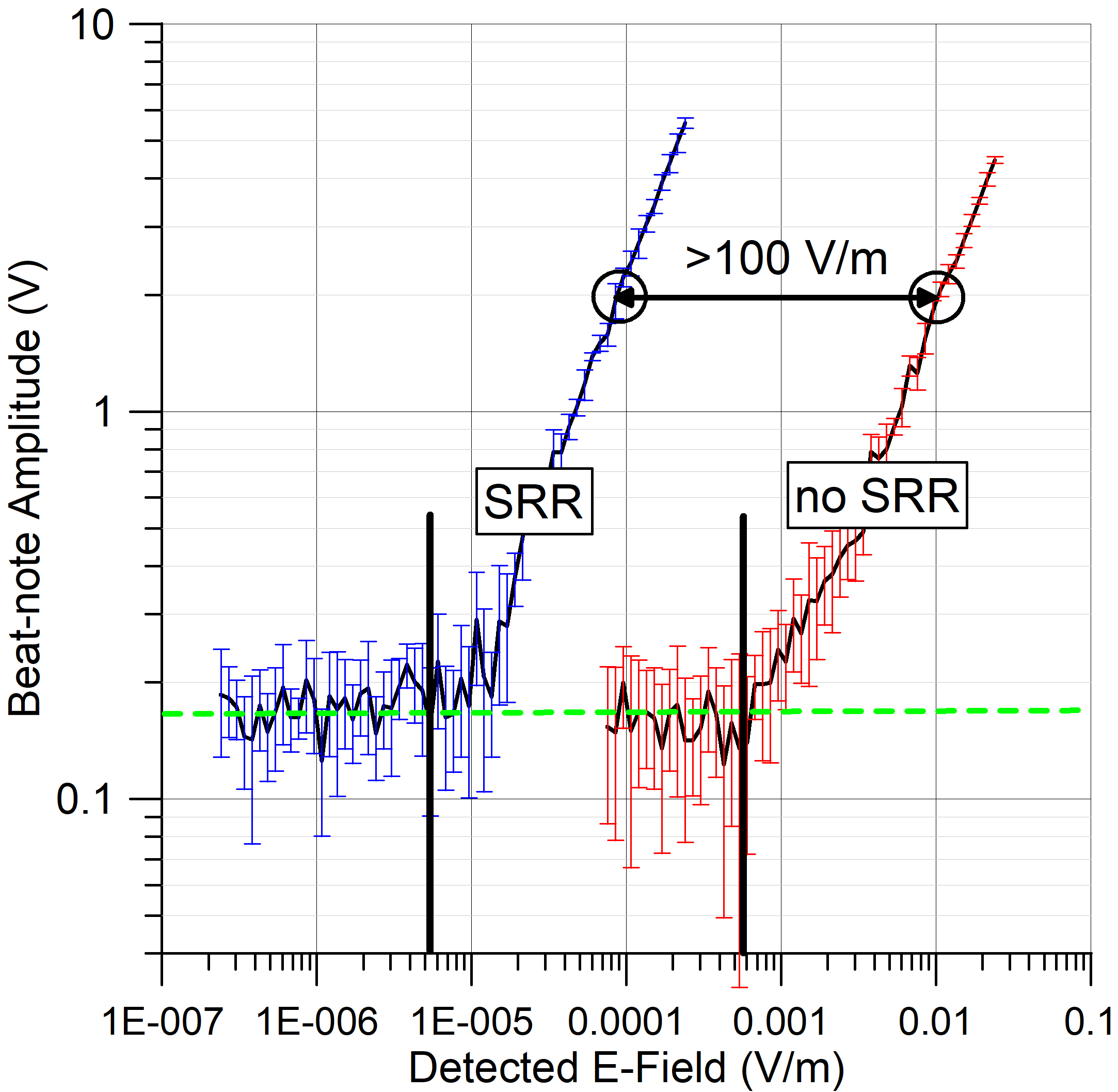}}\\
\vspace{-3mm}
\caption{Results from the Rydberg-atom mixer, plots of the output voltage of the lock-in amplifier as functions of the incident E-field.}
\label{mixer}
\end{figure}

The sensitivity of the standard EIT/AT is limited by the ability to resolve the AT splitting, which is typically limited to a few mV/cm$\sqrt{{\rm Hz}}$\,\cite{sed1, holl1, holl2, r15}. It has been shown that a heterodyne atom-based mixer approach can significantly improve the sensitivity of these types of Rydberg atom-based sensors \cite{gor3, jing1, Nik2021}. The details of the mixer techniques are found in Ref.\cite{gor3}. To implement this approach, we use a LO signal of 1.309~GHz+IF (where IF=10~kHz).  As discussed in Ref.\cite{gor3}, the beat-note amplitude from the atom-mixer indicates the ability to detect a weak E-field. 
Fig.~\ref{mixer} shows the output voltage from the lock-in amplifier (the beat note from the mixer) as a function of incident E-field strength (related to the output power of the SG) for the cases with and without the SRR. The error bars represent the standard deviation of 5 data sets. The two circles on the figure indicate that the two curves are shifted along the x-axis by two orders of magnitude, indicating an enhancement of 100. Without the SRR, we see in Fig.~\ref{mixer} that a E-field of 550~$\mu$V/m can be measured. Since we used a one-second averaging time for these results, this corresponds to a sensitivity of 550~$\mu$V/m$\sqrt{{\rm Hz}}$.  When the SRR is used, we see that an E-field of 5.5~$\mu$V/m can be measured, corresponding to a sensitivity of 5.5~$\mu$V/m$\sqrt{{\rm Hz}}$ (one-second averaging time). Comparing the two sets of results in Fig.~\ref{mixer} we see that the SRR case does show a field enhancement factor of 100. Consequently, the SRR can be used to substantially improve the sensitivity of these types of Rydberg-atom sensors.

We have demonstrated that simple SRRs can be used to enhance the incident E-field inside a vapor cell.  This enhancement facilitates a substantial improvement in detection of weak E-fields and improvement the sensitivity of Rydberg atom-based sensors. In fact, here we have demonstrated an enhancement factor of over 100. SRR designs with higher quality factors would increase this enhancement. Our SRR design was limited by the size of the vapor cell we had available (the 10.03~mm diameter of the cell), which limits the smallest gap we can use. A SRR with a smaller gap separation will allow for a larger field enhancement factor. Future work will include investigation of other resonators and cavities designs and other frequencies. We will also investigate incorporating these resonators (and cavities) with optical re-pumping\cite{Nik2021} and optical homodyne techniques\cite{kumar1} to further improve the sensitivity of these types of Rydberg atom-based sensors. Rydberg atom-based sensors are being designed for two distinct purposes. The first is to perform SI-traceable direct calibrated measurements, and the second is to design sensors and/or receivers where the absolute field values may not be required. The SRR is not a good choice when one is interested in a non-invasive sensor that would have minimal influence on the E-field (the SRR does perturb the field being measured). The SRR sensor for this purpose would require an additional calibration step for absolute field measurements as compared with a bare vapor cell. On the other hand, an SRR sensor is a good choice for receiver applications where weak field detection is desired. SRRs are a natural fit for integration with Rydberg atom sensors, as SRRs are sub-wavelength structures for enhancing a field strength and Rydberg atom sensors detect field strength on a sub-wavelength scale.

\newpage

\section*{Acknowledgements}
\vspace{-5mm}
\noindent This work was partially funded by the DARPA SAVaNT program and the NIST-on-a-Chip Program.

\section*{Data Availability Statement}
\vspace{-5mm}\noindent Data is available upon request.



\end{document}